\documentclass[aps,pre,epsf,superscriptaddress,amsmath,amssymb,amsfonts,article,showpacs]{revtex4-1}
\usepackage{amsfonts}
\usepackage{amsmath}

\usepackage{amssymb}
\usepackage{graphicx}
\usepackage{dcolumn}
\usepackage{times}
\usepackage{color}
\usepackage{appendix}
\begin{document}
  
\title{Stability and dynamics of dark-bright soliton bound states away from the integrable limit}

\author{G. C. Katsimiga}
\affiliation{Zentrum f\"ur Optische Quantentechnologien,
Universit\"at Hamburg, Luruper Chaussee 149, 22761 Hamburg, Germany}

\author{J. Stockhofe}
\affiliation{Zentrum f\"ur Optische Quantentechnologien,
Universit\"at Hamburg, Luruper Chaussee 149, 22761 Hamburg, Germany}

\author{P. G. Kevrekidis}
\affiliation{Department of Mathematics and Statistics, University of Massachusetts,
Amherst, MA 01003-4515, USA}

\author{P. Schmelcher}
\affiliation{Zentrum f\"ur Optische Quantentechnologien,
Universit\"at Hamburg, Luruper Chaussee 149, 22761 Hamburg, Germany}
 \affiliation{The Hamburg Centre for Ultrafast Imaging,
Luruper Chaussee 149, 22761 Hamburg, Germany}

\begin{abstract}
The existence, stability, and dynamics of bound pairs of symbiotic matter waves in the form of dark-bright soliton 
pairs in two-component mixtures of atomic Bose-Einstein condensates is investigated.
Motivated by the tunability of the atomic interactions in recent experiments,
we explore in detail the impact that changes in the interaction strengths have on these bound pairs
by considering significant deviations from the Manakov (integrable) limit.
It is found that dark-bright soliton pairs exist as stable configurations in a wide parametric window 
spanning both the miscible and the immiscible regime of interactions.
Outside this parameter interval two unstable regions are identified 
and are associated with a supercritical and a subcritical pitchfork bifurcation, respectively.  
Dynamical manifestation of these instabilities gives rise to 
a redistribution of the bright density between the dark solitons,
and also to symmetry-broken stationary states that are mass 
imbalanced (asymmetric) with respect to their bright soliton counterpart. 
The long-time dynamics of both the stable and the unstable balanced and imbalanced dark-bright soliton pairs is 
analyzed.
\end{abstract}

\maketitle

\section{Introduction}
After the experimental realization of Bose-Einstein condensates (BECs) in ultracold atoms, 
a plethora of studies has been devoted
to examining and understanding the coherent structures that arise in them~\cite{revip,emerge,frantznonlin,tomio}.
Among these structures, the formation, interactions and dynamics of matter wave dark~\cite{markus1,markus01,frantz} 
and bright solitons~\cite{emerge,borb0,borb1} have been a central focus
of research both from the experimental and from the theoretical side.  
Such nonlinear waves were experimentally generated in single-component BECs over a decade ago~\cite{g1,g2,g3,g4,carr2}. 
The nature of nonlinear matter waves that can be created in a BEC background, depends on the type of the interatomic
interactions. Namely, dark solitons can be created in BECs with atom-atom repulsion
resulting from a positive scattering length,
while bright solitons exist in single-component settings with attractive 
interatomic interactions resulting from a negative scattering length. 

In addition to the above single-component context, soliton states can 
arise also in multi-component settings. Such condensates have been created as 
mixtures of different spin states of $^{87}$Rb~\cite{bor1,bor2} and $^{23}$Na~\cite{bor3},
and triggered numerous theoretical studies  
involving soliton complexes. 
A prototypical example of the latter 
is a coupled dark-bright (DB) soliton state in a highly elongated (quasi-one-dimensional) condensate cloud, 
consisting of a dark soliton in one component and a bright soliton in the second component of a binary BEC featuring intra- and inter-species repulsion.
Since bright solitons are not self-sustained structures 
in repulsive (self-defocusing) media, DB solitons are often called symbiotic, 
that is the dark soliton can be thought of as acting as an
effective potential well
trapping the bright soliton ~\cite{DDB,pe5,va,beitia,carr}.  
Such symbiotic entities were first observed, and theoretically studied in the context of nonlinear 
optics~\cite{seg1,seg2,yuri1,christo,vdbysk1,vddyuri,ralak,dbysk2,shepkiv,parkshin}. 
However, their experimental realization in the atomic realm~\cite{hamburg} opened a new and highly controllable 
direction towards a deeper understanding of the dynamics and interactions of these states both with each other as well
as with external traps~\cite{BA,pe1,pe2,pe3,azu}.     

Additionally, current state-of-the art experiments offer the possibility of manipulating in a controllable
fashion the nonlinear interactions via the well-established technique of Feshbach 
resonances~\cite{fr1,fr10,fr100,fr2,fr20,chin,borisrh}.
This motivated recent theoretical activities  
where the static and dynamical properties 
of dark-bright symbiotic matter waves have been investigated on the level of mean field theory. 
Mathematically, tuning the interactions corresponds to deviating from the integrable (Manakov) limit \cite{Manakov} of 
the relevant nonlinear Schrödinger system, where nonintegrability is introduced 
when considering arbitrary nonlinearity (i.e., interaction) coefficients.
The latter nonintegrable setting forms also the main focus of the present effort.
In this context, despite the nonintegrability, analytical expressions of specific single-DB soliton states and lattices thereof have 
been obtained in~\cite{ef}.
Adding a parabolic trapping potential, it was revealed how the effective restoring force acting on the DB soliton depends on
the inter-atomic interactions~\cite{ef}, verifying that the particle-like nature~\cite{BA} of the symbiotic soliton is preserved.
In the same spirit, the dependence of the binding energy of a DB soliton 
on the inter-species interactions was found
analytically in~\cite{carr}, where moreover a proper bound 
on a phase imprinted in the bright soliton constituent 
was obtained (i.e. considering also moving single DB states), above which a breaking of the symbiotic entity was observed.
    
In our previous work~\cite{ljpp} the interactions between DB matter waves and the consequent 
formation of bound states for out-of-phase (anti-symmetric) bright soliton components has been studied. 
Based on a two-soliton ansatz of the hyperbolic type~\cite{pe3,ef},
the full
analytical expressions for the interaction energies between  
two DB solitons were obtained for arbitrary nonlinear coefficients, and in the absence of a confining potential.
Furthermore, the key intuition that 
repulsion mediated by the dark solitons at short distances, and 
attraction mediated by anti-symmetric bright solitons at longer 
distances, would be counterbalanced leading in turn to a bound state formation, 
has been enriched by taking into account the significant role of the cross-component interaction energy term.
The crucial dependence of the latter on the inter-species interaction coefficient has been analyzed.
It was shown that anti-symmetric stationary states
exist and remain robust for a wide parametric window of inter-species repulsions.
Importantly, an exponential instability of the anti-symmetric states was identified upon 
crossing a critical inter-species repulsion.
The latter was found to be associated with a subcritical 
pitchfork bifurcation, giving rise to asymmetric stationary states with mass imbalanced 
bright soliton counterparts.

In the present work, we extend the aforementioned findings of~\cite{ljpp} upon considering significant deviations from
the Manakov limit, towards both the immiscible (i.e., dominated by inter-species repulsion and thus phase separated in the 
ground state) regime,  
but also towards the largely unexplored miscible (i.e., dominated by intra-species repulsion) regime. 
To this end, we investigate the stability and dynamics 
of the anti-symmetric states, the so-called ``solitonic gluons''~\cite{seg2}, as well as 
the above-mentioned asymmetric modes covering both the miscible and immiscible parameter regime. 
In particular, the full excitation spectrum of these symbiotic states is explored in detail 
by means of a Bogolyubov-de Gennes linearization analysis~\cite{revip}.
It is found that for the asymmetric modes the linearization spectrum is strongly affected when crossing the immiscibility-
miscibility threshold,
which is in turn directly connected with a rapid change of their density profiles as observed already in~\cite{ljpp}.
As a next step, the stability and dynamics of the anti-symmetric DB pairs
is explored past the immiscibility-miscibility threshold. 
It is found that, in addition to the destabilization in the immiscible regime reported in~\cite{ljpp},
a second critical point occurs deep in the miscible regime, rendering the anti-symmetric state 
unstable once more. 
This instability scenario is found to be associated with a {\em supercritical} pitchfork bifurcation
giving rise to another family of mass imbalanced (asymmetric) symbiotic structures which are found to be stable. As the 
inter-species repulsion is decreased,
the miscible character of this regime alters the bright soliton component, resulting into asymmetric pairs with the bright 
solitons ``living'' on top of a finite background.
Comparing and contrasting the instability mechanisms in the different parameter regimes,      
the long-time evolution of the asymmetric and anti-symmetric states 
is performed numerically at different interaction ratios.

The presentation is structured as follows. In Sec. 2 a description of the theoretical model and 
prior results regarding the existence of the anti-symmetric DB soliton pairs is provided. 
Furthermore, we briefly comment on the methods to be used 
for the numerical analysis of our findings. 
Sec. 3 contains the results regarding both the stability and the dynamics 
of asymmetric and anti-symmetric DB soliton pairs beyond the integrable limit. 
Finally, in Sec. 4 we summarize our findings and discuss 
future perspectives. 
\section{Setup and prior background}

\subsection{Model and theoretical considerations}
The model of interest is a system of a two-component BEC strongly elongated along the $x$-direction,
subject to a tight transverse harmonic trap of frequency $\omega_\perp$. 
Such a mixture can e.g be composed of two different
hyperfine states of the same alkali isotope, like $^{87}$Rb.
Within mean field theory and after integrating out the frozen  transverse degrees of freedom, 
this mixture is described by the following two coupled (1+1)-dimensional Gross-Pitaevskii equations (GPEs) 
\cite{stringari,siambook}:
\begin{eqnarray}
i\hbar \partial_t \psi_j =
\left( -\frac{\hbar^2}{2m} \partial_{x}^2 -\mu_j + \sum_{k=1}^2
g_{jk} |\psi_k|^2\right)\!\psi_j. \quad \quad (j=1,2)
\label{model}
\end{eqnarray}
In the above equation, $\psi_j(x,t)$ denote the mean-field wave functions of the 
two components normalized to the
numbers of atoms $N_j = \int_{-\infty}^{+\infty} |\psi_j|^2 dx$, while $m$ and $\mu_j$ 
are the atomic mass (identical for both components) and chemical potentials, respectively. 
The effective one-dimensional coupling constants are given by $g_{jk}=2\hbar\omega_{\perp} a_{jk}$, 
where $a_{jk}$ denote the three $s$-wave scattering lengths that account for collisions between atoms belonging to 
the same ($a_{jj}$) or different ($a_{12}=a_{21}$) species.
We restrict our considerations
to purely repulsive interactions, i.e. all $g_{jk}>0$, 
and consider an idealized homogeneous setting with no longitudinal trapping potential along the $x$-axis.

By measuring densities {$|\psi_j|^2$}, 
length, time and energy in
units of $2a_{11}$, $a_{\perp} = \sqrt{\hbar/\left(m \omega_{\perp}\right)}$, 
$\omega_{\perp}^{-1}$, $\hbar\omega_{\perp}$ respectively, and in a second step rescaling space-time coordinates
as $t \rightarrow 
\mu_1 t$, $x \rightarrow {\sqrt{\mu_1}}x$,
and the densities $|\psi_{1,2}|^2 \rightarrow \mu_{1}^{-1} |\psi_{1,2}|^2$, 
the system of Eqs.~(\ref{model}) can be written in the following dimensionless form: 
\begin{eqnarray}
i \partial_t \psi_d + \frac{1}{2} \partial_{x}^2\psi_d  
-(|\psi_d |^2 + g_{12}|\psi_b |^2 -1) \psi_d &=& 0,
\label{deq11}
\\
i \partial_t \psi_b +\frac{1}{2} \partial_{x}^2\psi_b  
-(g_{12}|\psi_d |^2 + g_{22}|\psi_b |^2- \mu) \psi_b &=& 0.
\label{deq21}
\end{eqnarray}
In the above system of equations we slightly changed the notation, using
$\psi_1 \equiv  \psi_d$ and $\psi_2 \equiv  \psi_b$, 
indicating this way that the component $1$ ($2$) will be 
supporting dark (bright) solitons. Furthermore, $\mu\equiv\mu_2/\mu_1$ is the rescaled chemical potential,
while the interaction coefficients are normalized to the scattering 
length $a_{11}$, i.e. $g_{12}\equiv g_{12}/g_{11}$, 
and $g_{22}\equiv g_{22}/g_{11}$. 
The system of Eqs.~(\ref{deq11})-(\ref{deq21}) conserves the total energy, $E$, the rescaled total number of atoms in each 
component ($N_d$ and $N_b$, respectively) and the rescaled total number of atoms $N$, where:
\begin{eqnarray}
E &=& \frac{1}{2}\int_{-\infty}^{+\infty} \Big[|\partial_{x} \psi_d|^2+|\partial_{x} \psi_b |^2+(|\psi_d |^2-1)^2
+g_{22}|\psi_b |^4-2\mu |\psi_b |^2 + 2g_{12} |\psi_d |^2 |\psi_b |^2\Big] dx, 
\label{energy}  \\
N&=&N_d+N_b=\sum_{i=d,b}\int^{\infty}_{-\infty} dx|\psi_i |^2.
\label{Ntot}
\end{eqnarray}
\subsection{Interactions of symbiotic matter waves beyond the Manakov limit}
In the special case where the nonlinear coefficients are all equal to each other, i.e. $g_{12}=g_{22}=1$, Eqs.~(\ref{deq11})-(\ref{deq21}) correspond to the integrable Manakov 
model~\cite{Manakov}. In such a case, it particularly admits exact single-DB-soliton solutions of the form~\cite{BA}:
\begin{eqnarray}
\!\!\!\!\!\!
\psi_d(x,t)&=&\cos\phi\tanh\left[D(x-x_0(t))\right]+i\sin\phi,
\label{dark}\\
\psi_b(x,t)&=&\eta {\rm sech}\left[D(x-x_0(t))\right] 
\times \exp\left[ikx+i\theta(t)+i(\mu-1)t\right],
\label{bright}
\end{eqnarray}
subject to the boundary conditions $|\psi_d|^2\rightarrow 1$, and $|\psi_b|^2\rightarrow 0$ for $|x|\rightarrow  \infty$.
In the aforementioned solutions, $\phi$ is the so-called soliton's phase 
angle which fixes the ``grayness'' of the dark soliton, while $\eta$ denotes the amplitude of the bright 
component. Furthermore, $x_0(t)$ and $D$ correspond to the soliton's center 
and inverse width respectively, while $k=D \tan \phi$ is the wave-number of the bright soliton, associated with the
speed $\dot x_0$ of the DB soliton, and $\theta(t)$ is its phase.

While such a general exact expression can only be obtained in the Manakov limit, 
one can utilize an approximate ansatz 
based on the expressions given in Eqs.~(\ref{dark})-(\ref{bright}) and depart
from the integrable limit.
Such a method was employed in our very recent work of Ref.~\cite{ljpp} in order to analyze effective interactions between  
symbiotic matter waves in the general case of different inter-atomic repulsion strengths within each and between the species. 
Our aim in what follows is to briefly comment on the key results obtained there,
to be connected with the numerical findings that will be presented below.
In particular, a pair of two equal-amplitude dark-bright solitons travelling in opposite directions 
was considered having the approximate form:
\begin{eqnarray}
\psi_d(x,t)
&=& \left(\cos\phi\tanh X_{-}+i\sin\phi \right)
\times 
\left(\cos\phi\tanh X_{+}-i\sin\phi\right),
\label{2dbd}
\\[1.0ex]
\psi_b(x,t)
&=& \eta\, {\rm sech} X_{-}\, {\rm e}^{i\left[kx+\theta(t)
+(\mu-1)t
\right]}+
\eta\, {\rm sech} X_{+}\, {\rm e}^{i\left[-kx+\theta(t)
+(\mu-1)t
\right]}\,
{\rm e}^{i\Delta\theta}.
\label{2dbb}
\end{eqnarray}
In these expressions $X_{\pm} = D\left(x \pm x_0(t)\right)$ where $2x_0(t)$ is the distance between the two DB solitons, while 
$\Delta\theta$ is the 
relative phase between the bright solitons. 
If $\Delta\theta=0$, the bright solitons are in-phase (IP), while if $\Delta\theta=\pi$,
the bright solitons are out-of-phase (OP) or anti-symmetric.
Within a Hamiltonian variational approach~\cite{pe3,ef,frantz,frantznonlin} 
the approximate ansatz of 
Eqs.~(\ref{2dbd})-(\ref{2dbb}) is substituted into the total energy of the system 
given by Eq.~(\ref{energy}), and the relevant integrations are performed 
under the assumption that the soliton velocity is sufficiently small ($\phi \approx 0$, $k\approx 0$). 
The total energy then can be decomposed as $E=2E_1+E_{dd}+E_{bb}+E_{db}$,
where $E_1$ corresponds to the energy of a single DB soliton, contributing
twice to the total energy, while the remaining three terms, namely $E_{dd}$, $E_{bb}$ and $E_{db}$, 
account for the interaction between: the two dark solitons,  
the two bright solitons, and the cross-interaction  
of a dark soliton in the first component 
with the bright soliton of the second component, respectively.
Analytical expressions for each of these energy contributions as a function of the DB soliton distance $x_0$ and for arbitrary 
nonlinear coefficients can be found in~\cite{ljpp},
together with simplified asymptotic expansions valid for large distances, $D x_0 \gg 1$.

\begin{figure}[tbp]
\centering
\includegraphics[scale=0.35]{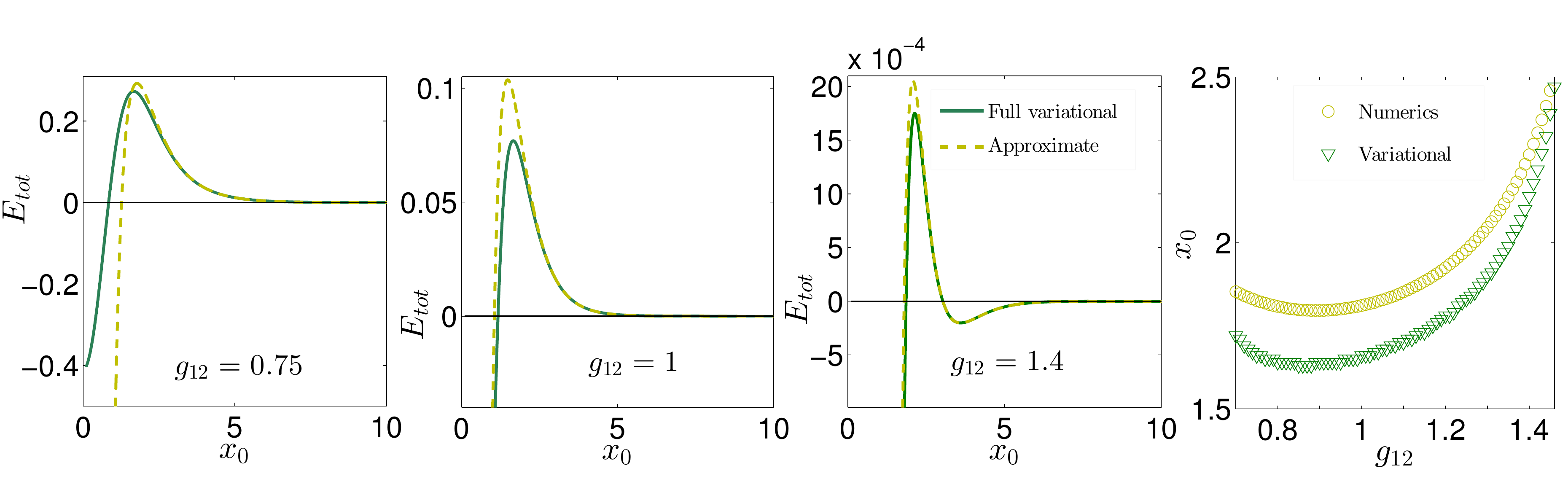}
\caption{Variational estimate of the total interaction energy $E_{tot}$ between two  
OP dark-bright solitons. From left to right 
in panels (\textbf{a})-(\textbf{c}) the value of the inter-species interaction coefficient is increased from $g_{12}=0.75$ to 
$g_{12}=1.4$, while $g_{22}=0.95$ and $\mu=2/3$ throughout. In all cases both the full and the asymptotic (valid for large $x_0$) 
expressions are shown with solid green 
and dashed yellow lines, respectively. 
The equilibrium separation $x_0$ of the stationary DB soliton pair as a function of $g_{12}$ is 
shown in panel (\textbf{d}). 
In this latter panel circles in yellow correspond to the equilibrium distance obtained by numerically solving 
the GPE system of Eqs.~(\ref{deq11})-(\ref{deq21}), while triangles in green denote the variational prediction of the 
equilibrium, i.e. the respective maximum of $E_{tot}$.}
\label{fig1}
\end{figure}  

Figure~\ref{fig1} illustrates our key findings regarding the interaction energy of two out-of-phase DB solitons as a function of their 
distance.
Here and throughout this work the chemical potential and the intra-species interaction 
coefficients are fixed to $\mu=2/3$ and $g_{22}=0.95$~\cite{halll,opanchuk}, respectively, 
while the interatomic interaction coefficient $g_{12}$ is left to widely vary. 
In panels (a)-(c) both the full and the asymptotic
forms of the analytically obtained expressions for the total interaction energy, $E_{tot}=E_{dd}+E_{bb}+E_{db}$, 
are shown with solid green and dashed yellow lines, respectively. 
As it can be seen in all cases, the two results coincide at large distances as they should.
Strikingly, in all three cases a pronounced local energy maximum is identified. The identification of this extremum suggests, if we 
further take into account the effective negative mass of the DB soliton~\cite{frantz},   
the existence of an effective \emph{stable} (with respect to variations in $x_0$) fixed point,
thus a bound state for the two OP dark-bright solitons. 
Such an anti-symmetric two-DB soliton bound state can indeed be identified also in the full Gross-Pitaevskii system (for the numerical 
methods used we refer the reader to the following section).
The predictive strength of the variational approximation is directly checked by comparing 
with the full numerical computation of the respective equilibria $x_0$ on the level of Eqs.~(\ref{deq11})-(\ref{deq21}), and the 
outcome is illustrated in panel (d), showing good agreement.

It is worth mentioning at this point  
that in order to evaluate the energy functional we need as input the width $D$ and amplitude $\eta$ occuring in Eqs.~(\ref{2dbd})-(\ref{2dbb}).
These are obtained by numerically identifying the exact single-DB state at the given $g_{12}$ and fitting to it
the profile known from the integrable limit, i.e. Eqs.~(\ref{dark})-(\ref{bright}). 
There are thus two main sources of error of this scheme, namely (i) the imperfect fit of the $tanh$-$sech$ profile (strictly valid only in the integrable limit) to the single DB soliton mode
and (ii) the limited accuracy of the two-DB soliton ansatz of Eqs.~(\ref{2dbd})-(\ref{2dbb}) especially at small separations $x_0$.
For details regarding this and also a discussion of the case of in-phase bright solitons we refer the interested reader to Ref.~\cite{ljpp}.
We note also that as $g_{12}$ increases (while $\mu$ and $g_{22}$ are kept fixed), the norm $N_b$ of the bright species is increasingly suppressed
and eventually vanishes \cite{ljpp}, beyond which point there is no bound two-soliton state anymore, since the presence of the bright component is a necessary
ingredient for holding the dark solitons together \cite{seg2,pe3}. This suppression of the bright norm is compatible with the overall decrease of the total energies from panel (a)  
towards (b) and (c) in Figure~\ref{fig1}.

Having variationally predicted and numerically confirmed the existence of anti-symmetric stationary dark-bright soliton bound states 
for a large interval of values of the inter-species interaction parameter $g_{12}$ (typically, we study $0.75 \leq g_{12} \leq1.5$ in the present work), our main goal in the following is
to address their stability and (where applicable) decay dynamics. 

\subsection{Numerical methods}
In this section we briefly comment on the numerical methods to be used 
so as to obtain stationary symbiotic states consisting of two dark-bright solitons 
and to determine their stability and time evolution. 
In the numerical computations that follow, we initially obtain stationary solutions of the system of Eqs.~(\ref{deq11})-(\ref{deq21})
in the form of $\psi_d(x,t)=u_d(x)$ and $\psi_b(x,t)=u_b(x)$ by means of a fixed-point numerical 
iteration scheme~\cite{kelley}. 
The linear stability of the latter is adressed by using the Bogolyubov-de Gennes (BdG) analysis~\cite{siambook,frantz}.
In particular to assess the stability of the obtained fixed points, we substitute the following ansatz
into Eqs.~(\ref{deq11})-(\ref{deq21}):
\begin{eqnarray}
  \psi_d(x,t) &=& u_d(x) +\epsilon \left(a(x) e^{-i \omega t} + b^{\star}(x) e^{i \omega^{\star} t}
  \right),
  \label{bdg1}
  \\
 \psi_b (x,t)&=& u_b(x) + \epsilon \left(c(x) e^{-i \omega t} + d^{\star}(x) e^{i \omega^{\star} t}
  \right). 
  \label{bdg2}
\end{eqnarray}
In the above equations the asterisk denotes complex conjugation while $\epsilon$ is the amplitude of 
infinitesimal perturbations. The resulting system of equations is then linearized, 
by keeping only terms of order $\mathcal{O}(\epsilon)$, 
and the eigenvalue problem for the eigenfrequencies $\omega$ and
eigenmodes $(a(x),b(x),c(x),d(x))^T$ is solved numerically. 
Note that an instability occurs if modes with purely imaginary or complex
eigenfrequencies are identified~\cite{revip}. Since the linearization spectrum is invariant under $\omega \rightarrow -\omega$
and $\omega \rightarrow \omega^\star$, only results in the first quadrant of the complex plane will be shown below.
For the simulation of the time evolution based on Eqs.~(\ref{deq11})-(\ref{deq21}), a fourth order Runge-Kutta algorithm is 
employed.
In all cases, the numerical computations are performed on a spatial grid in the presence of an almost hard-wall super-Gaussian 
potential \cite{ljpp} that is chosen wide enough for boundary effects to be negligible on small and intermediate time scales.

\section{Numerical results}
\begin{figure}[tbp]
\centering
\includegraphics[scale=0.4]{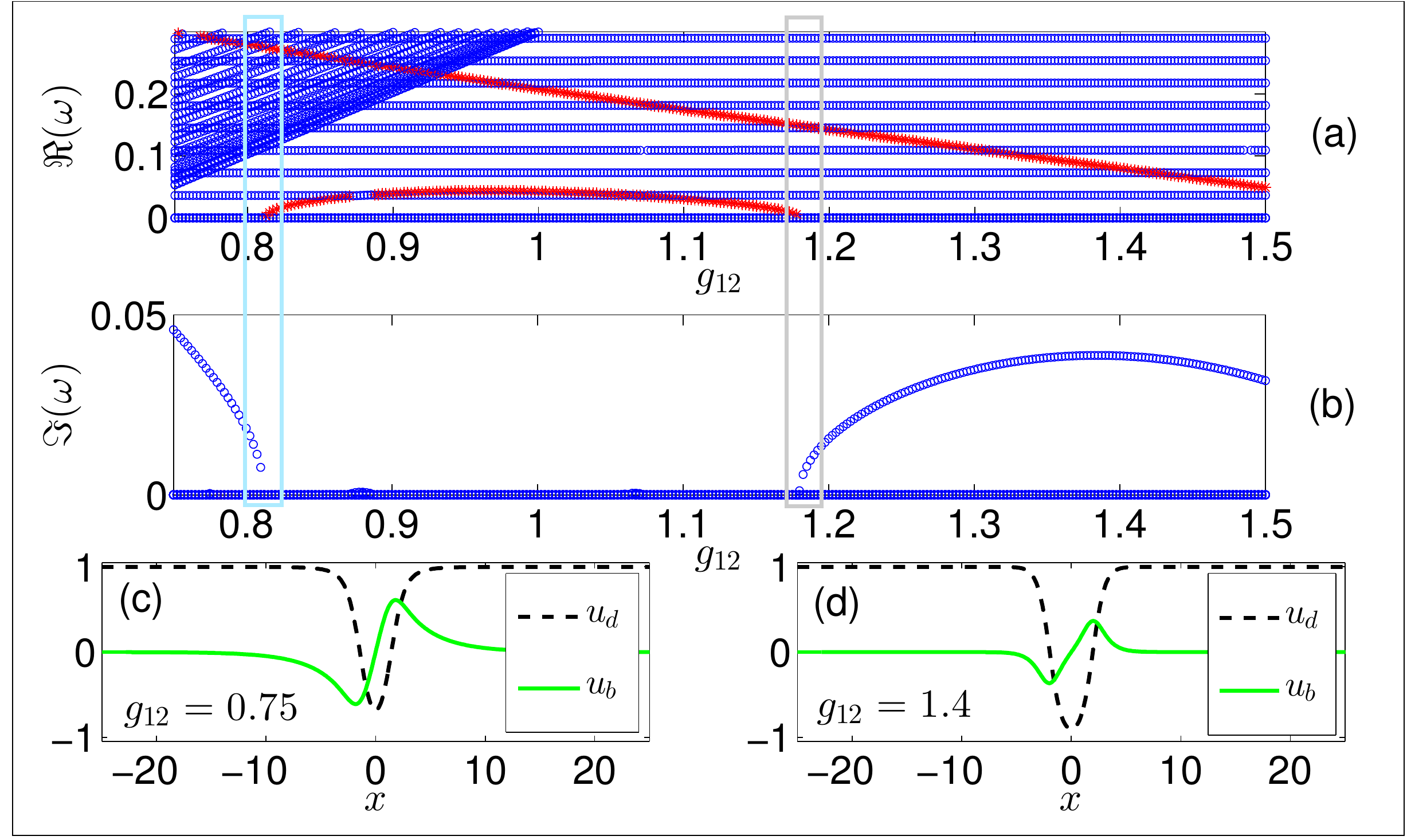}
\caption{Full BdG spectrum of anti-symmetric two-DB soliton states, 
upon varying the inter-species interaction coefficient 
$g_{12}$ within the interval $0.75 \leq g_{12} \leq 1.5$. (\textbf{a}) Real part, $\Re(\omega)$, of the eigenfrequencies $\omega$ as a 
function of $g_{12}$. The corresponding imaginary part, $\Im(\omega)$, is shown in panel (\textbf{b}). 
Upon increasing $g_{12}$ there exist two critical values, $g^{(2)}_{12_{cr}}=0.81$, and $g^{(1)}_{12_{cr}}=1.18$, indicated with 
light blue and gray boxes respectively, below and above which the anti-symmetric 
branch destabilizes. The trajectories of the two anomalous modes (see text) appearing in the spectrum 
are shown with red stars. (\textbf{c})-(\textbf{d})  
Profiles of the anti-symmetric states for $g_{12}=0.75$, and $g_{12}=1.4$, i.e. deep in the miscible and immiscible regime,
respectively.}
\label{fig2}
\end{figure}   
Having verified that stationary anti-symmetric pairs of symbiotic matter waves
can be found in a wide range of values of the inter-species repulsion coefficient $g_{12}$,
a natural next step is to consider the fate of these solutions under small perturbations,
providing information on their stability in the different regions of existence.
The latter is explored by using the BdG linearization
analysis discussed in Section~2.3.
For the presentation of our results, we will distinguish the miscible ($g_{12} < g_{12_{th}}$) and immiscible  ($g_{12} > 
g_{12_{th}}$) regimes,
separated by the miscibility-immiscibility threshold~\cite{aochui} $g_{12_{th}} = \sqrt{g_{22}}=0.975$ for our choice 
of $g_{22}=0.95$.    
In Figure~\ref{fig2}, 
the BdG spectrum of the anti-symmetric stationary two dark-bright soliton
states is shown as a function of $g_{12}$. 
In particular, both the real $\left(\Re(\omega)\right)$ and the 
imaginary parts $(\Im(\omega))$ of the eigenfrequencies $\omega$ 
are depicted in the top (a) and middle (b) panels respectively.
Panels (c) and (d) depict profiles of the dark and bright wave functions at selected values of $g_{12}$ 
in the miscible and immiscible regime, respectively.
Two general comments can be made before examining in detail the excitation spectrum.
The most significant one is that within the background spectrum denoted with blue circles 
there exist two distinguished modes. The trajectories 
of the latter are illustrated with red stars.
These modes are the so-called anomalous modes since they posses a so-called negative Krein
signature $K$~\cite{skryabin}, which for the two-component system considered herein is defined as:
\begin{equation}
K=\omega\int\left(|a|^2-|b|^2+|c|^2-|d|^2\right)dx.
\end{equation}
The sign of this quantity is a topological property associated with
the excited nature of this state and the eigenvectors of such
anomalous modes result in
a variation of the solitary waves (as opposed to a variation of
the system's background).
Furthermore, we note in passing that each continuous symmetry of the system corresponds to a pair of zero modes, 
$\omega=0$, in the BdG spectrum. Thus we expect three pairs of such modes related respectively with 
the conservation of the particle number (or the $U(1)$ gauge invariance) in each of the two components, and with the 
translation invariance due to the absence of a confining potential. This is confirmed in the numerical data.

Just by inspecting the trajectories of the two anomalous modes that appear in the spectrum, 
their very different behavior becomes apparent.
As it is observed in panel (a) of Figure~\ref{fig2}, 
the higher frequency mode decreases almost monotonically upon increasing $g_{12}$.
We were able not only to identify this mode but also to relate it with an out-of-phase vibration of the bound DB soliton pair 
around its equilibrium distance. 
The latter can be done by adding the corresponding eigenvector 
for some value of the inter-species interaction coefficient to the relevant stationary 
anti-symmetric state (results not shown here). 
Next, and also in the same panel of Figure~\ref{fig2}, let us closely follow the trajectory of the lower-lying anomalous mode. 
The trajectory of this mode is more complicated than the former one.
In particular, starting from the aforementioned reference point, $g_{12_{th}} \approx 1$, and increasing the interspecies 
interactions towards the immiscible regime, i.e. for $g_{12}>g_{12_{th}}$, it can be seen that there exists a critical point 
$g^{(1)}_{12_{cr}}=1.18$ where this mode destabilizes.
This destabilization corresponds to an eigenvalue zero crossing and signals the instability of the
anti-symmetric configuration deep in the immiscible regime. Notice the non-zero imaginary part that appears past this critical 
point shown in panel (b) of Figure~\ref{fig2}.   
The existence and destabilization of this mode was found to be related with a symmetry breaking 
of the bright soliton component, being linked, in turn, to other stationary states that are mass imbalanced with respect to 
their bright soliton counterpart.
The identification of these asymmetric states which exist below $g^{(1)}_{12_{cr}}$ and collide with the anti-symmetric branch
at $g^{(1)}_{12_{cr}}$ in a subcritical pitchfork bifurcation was established in~\cite{ljpp}. 
The following paragraphs will be devoted to a discussion of these asymmetric two-DB solutions, 
before we then return to an analysis of the BdG spectrum of the anti-symmetric pair, Figure~\ref{fig2}, at small values of $g_{12}$.
\begin{figure}[tbp]
\centering
\includegraphics[scale=0.42]{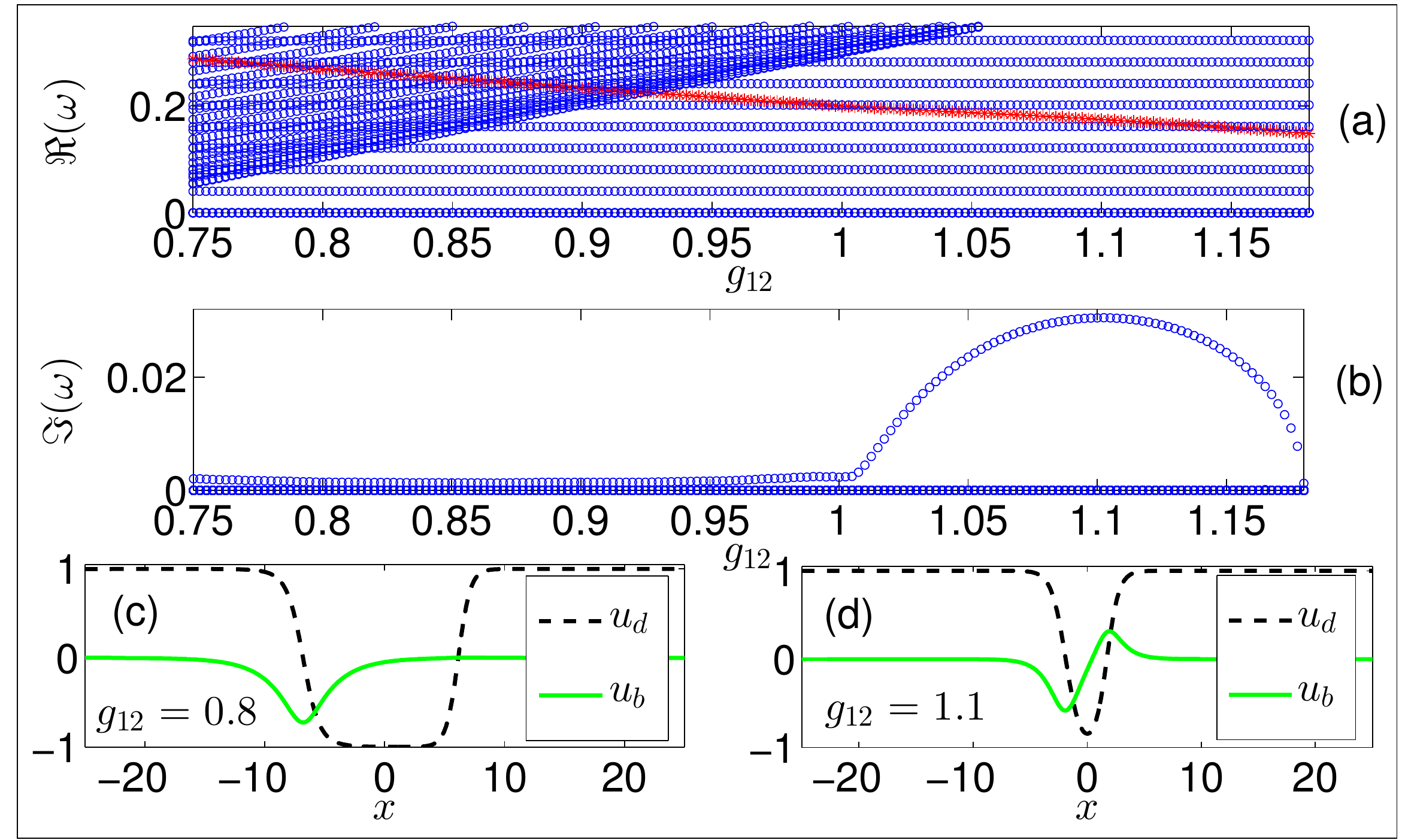}
\caption{Same as Figure~\ref{fig2} but for the asymmetric two-DB soliton states.
  Notice in this case the significant modification of the stability
  properties (i.e., a nearly vanishing imaginary part of the eigenfrequency
  in panel (b)), as the miscibility-immiscibility threshold
  $g_{{12}_{th}} \approx 1$ is crossed.}
\label{fig3}
\end{figure}   

As mentioned above, the asymmetric and the anti-symmetric two-DB states coincide at the bifurcation point $g^{(1)}_{12_{cr}}$.
Below this critical $g_{12}$, the anti-symmetric state is stable and there are two symmetry-broken asymmetric solutions which 
are unstable (as is characteristic of a subcritical pitchfork bifurcation). Indeed, these asymmetric branches can themselves 
be continued towards much smaller values of $g_{12}$ and exist both in the immiscible and the miscible regime, 
see the profiles shown in Figure~\ref{fig3} (c) and (d). 
In both panels, dashed black lines denote the wavefunction of the dark soliton component,
while solid green lines depict the corresponding bright soliton counterpart.
To gain further insight regarding the nature of the instability of these  
asymmetric states, their full BdG spectrum is illustrated in Figure~\ref{fig3} (a) and (b).
Once more, both the real $\Re(\omega)$ and the imaginary $\Im(\omega)$ parts of the eigenfrequencies $\omega$ are shown
as a function of the interspecies interaction. 
As expected from the above discussion, only one of the anomalous modes appears in the excitation spectrum of these mass 
imbalanced states and is depicted with stars in red.
Replacing the second anomalous mode, there is now throughout a purely imaginary frequency signaling the instability of the 
asymmetric branch.  
Remarkably, as the immiscibility-miscibility threshold is crossed, 
the growth rate of the instability is drastically decreased, rendering these states only weakly unstable for $g_{12} < 
g_{{12}_{th}}$. 
This latter observation is in close contact with the change in the spatial character
of these asymmetric states, and also in line with our previous findings~\cite{ljpp}. 
Namely, as the interspecies interactions decrease towards the miscible region, the asymmetric states change gradually from 
only weakly asymmetric (i.e. weakly mass imbalanced with respect to their bright amplitudes), to maximally asymmetric (i.e.
almost purely dark/dark-bright bound states).
The difference is clearly seen in the profiles provided in panels (c) and (d) of Figure~\ref{fig3}.

\begin{figure}[tbp]
\centering
\includegraphics[scale=0.4]{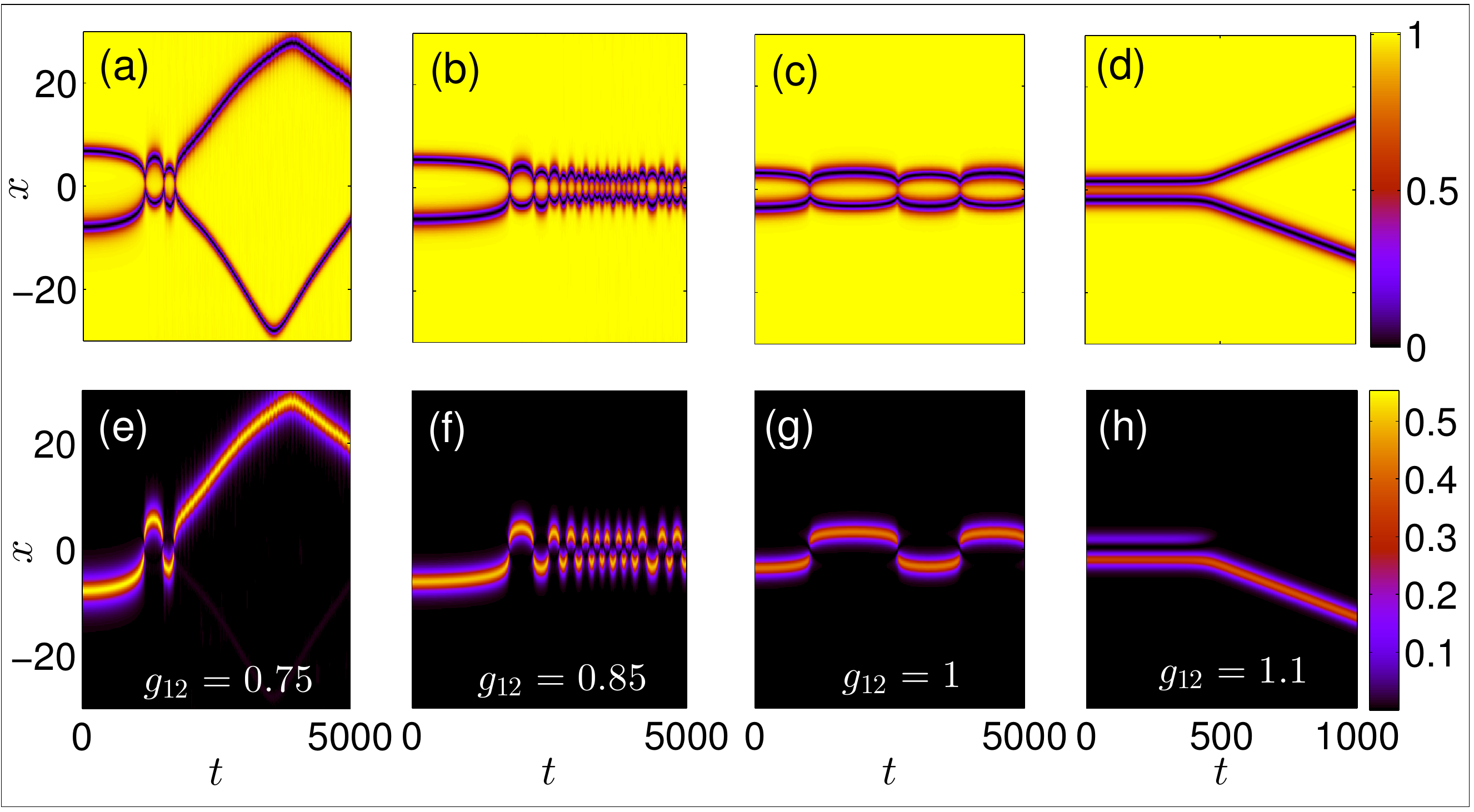}
\caption{Space-time evolution of the unstable asymmetric two-DB soliton state for different values of the inter-species 
interaction coefficient $g_{12}$. 
From left to right the interaction is increased from $g_{12}=0.75$ to 
$g_{12}=1.1$, while panels (\textbf{a})-(\textbf{d}) [(\textbf{e})-(\textbf{h})] 
correspond to the densities $|\psi_d|^2$ ($|\psi_b|^2$) 
of the dark (bright) component.}
\label{fig4}
\end{figure}

Having identified the unstable linearization eigenmodes, we now turn to the associated decay dynamics.   
In particular, Figure~\ref{fig4} shows the long-time dynamics
of the above-obtained stationary asymmetric states 
for different values of the nonlinear coefficient $g_{12}$. 
Panels (a)-(d) correspond to the dark soliton 
component while the respective bright wave functions are depicted in panels (e)-(h).
For $g_{12}<1$ shown in panels (a)-(c) and (e)-(g) respectively, it is observed
that as a result of the instability, the solitons move towards each other 
forming breathing-like structures. 
This is accompanied by a redistribution of the bright ``filling'' component between the two dark solitons, in the form 
of almost complete tunneling back and forth.
For parameter values slightly above $g_{12_{th}}$, as e.g. the one depicted in panels (c) and (g),
a strong beating phenomenon is observed, clearly evident in the dark soliton component, with the solitons oscillating around a 
fixed distance from each other forming an almost stationary breather that persists till the end of the propagation.
Below the above-mentioned threshold and towards the miscible region the picture becomes progressively more dramatic.        
The beating gets much more pronounced with the solitons experiencing more frequent collisions as shown in panels (b) and (f). 
Finally, when entering even deeper in the miscible side illustrated in panels (a) and (e), eventually the bound pair fully 
splits into an essentially empty dark and a dark-bright soliton that are released
(i.e., are no longer bound by each other)
and propagate towards the outer parts of the 
simulation domain (where they are ultimately reflected by the boundaries).     
However, for $g_{12}>1$, i.e. upon increasing $g_{12}$ towards the immiscible side depicted in panels (d) and (h) of 
Figure~\ref{fig4}, a rather different picture is painted by the symbiotic entities when compared to the asymmetric states  
presented above. In particular, slightly after the beginning of the propagation, where the asymmetric entity looked quite 
robust, a dramatic redistribution of the bright soliton's mass occurs. The latter results in a strong repulsion between
the single DB soliton formed and the almost empty dark one leading in turn to their subsequent separation.
Note that a similar decaying mechanism was also observed in~\cite{ljpp} but for the unstable anti-symmetric states 
(see also later on in the text).   

We have now characterized the branch of asymmetric two-DB soliton modes, which were seen to 
bifurcate from the anti-symmetric two-DB soliton branch in a subcritical pitchfork bifurcation in the immiscible regime,
at $g^{(1)}_{12_{cr}}=1.18$.
Let us now return to the anti-symmetric branch itself and study its fate in the
miscible regime.
This regime is also covered in the full excitation spectrum depicted in Figure~\ref{fig2}, with a typical profile of the state 
being shown in panel (c).
Departing from $g_{12}\approx1$ towards the miscible regime, and in particular by following once again the lower-lying 
anomalous BdG 
mode, we observe that as $g_{12}$ decreases a second critical point exists deep within the miscible side.
The eigenvalue zero crossing occurs at $g^{(2)}_{12_{cr}}=0.81$, 
and is indicated by the solid light blue box. 
We note here, that this latter critical point was not discussed in our previous work~\cite{ljpp}, 
where only larger values of $g_{12}$ were studied.
The destabilization of the aforementioned mode suggests the existence of a second pitchfork bifurcation. 
This is, once again, related with a symmetry breaking of the bright soliton component resulting
into mass imbalanced (i.e., asymmetric) two DB states.
\begin{figure}[tbp]
\centering
\includegraphics[scale=0.35]{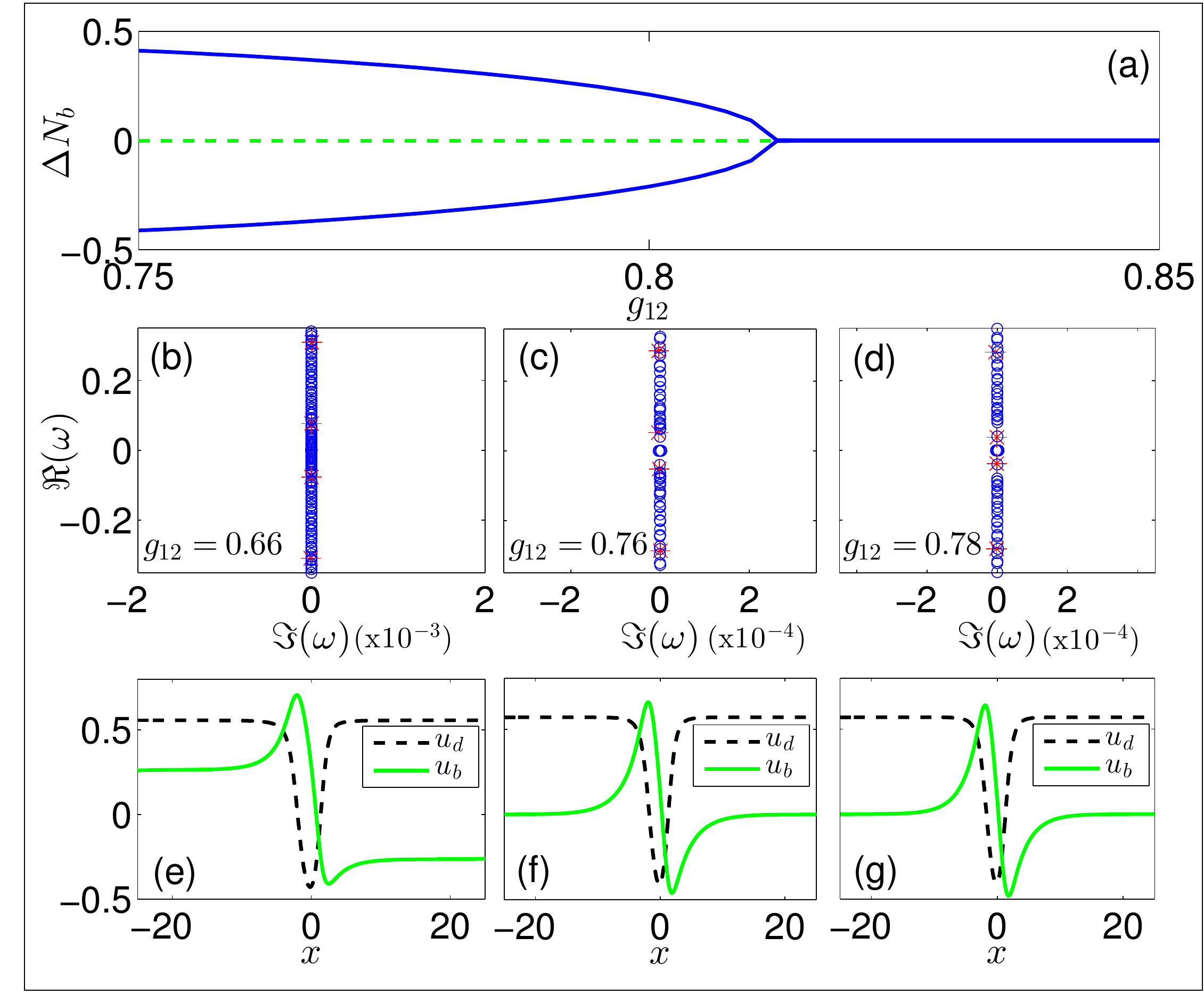}
\includegraphics[scale=0.35]{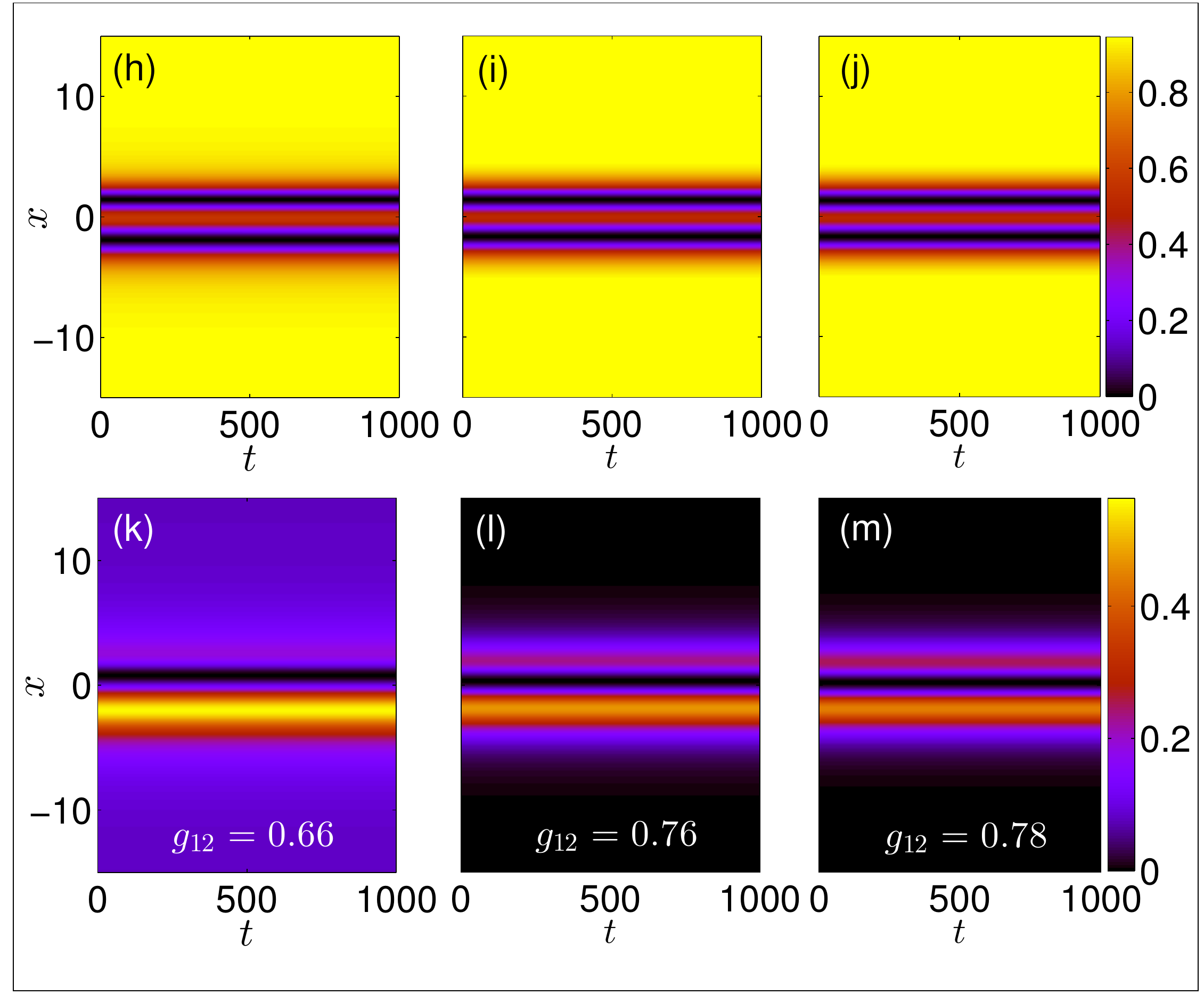}
\caption{(\textbf{a}) Bifurcation diagram obtained by measuring the relative bright imbalance $\Delta N_b$ 
as a function of the inter-species 
interaction coefficient $g_{12}$. The stable anti-symmetric and asymmetric branches 
are denoted with solid blue lines, while the unstable anti-symmetric one is shown with dashed green line.
Notice that the asymmetric branches exist before the critical point, $g^{(2)}_{12_{cr}}=0.81$, 
but are stable verifying the supercritical nature of the bifurcation. (\textbf{b})-(\textbf{d}) BdG spectra at three different  
values of $g_{12}$ showcasing the stability of the asymmetric states. 
In all cases the anomalous modes are illustrated with red stars.
The associated stationary DB profiles are depicted in 
panels (\textbf{e})-(\textbf{g}).  
(\textbf{h})-(\textbf{m}) Spatio-temporal evolution of the stationary asymmetric states
shown in panels (\textbf{e})-(\textbf{g}) showing the dynamical stability of these symbiotic structures. Panels 
(\textbf{h})-(\textbf{j}) [(\textbf{k})-(\textbf{m})] correspond to the dark (bright) soliton component.}
\label{fig5}
\end{figure}   

Indeed, we were able to identify these 
{\em new} pairs. In contrast to the previous instability scenario, 
these mass imbalanced states exist past the critical point but are {\em stable}, i.e. the pitchfork  
deep in the miscible domain is found to be {\em supercritical}.
The corresponding bifurcation diagram is shown in panel (a)  
of Figure~\ref{fig5}. 
In order to obtain this diagram we measure
the relative bright imbalance defined as $\Delta N_b\equiv\left(\int_{-\infty}^0 |u_b|^2 dx-\int^{\infty}_0 |
u_b|^2 dx\right)/\int^{\infty}_{-\infty}|u_b|^2dx$ upon varying $g_{12}$. Notice that four branches are identified,
i.e. three stable ones consisting of two asymmetric and an anti-symmetric branch all denoted by solid blue lines, and one 
unstable anti-symmetric branch illustrated with dashed green line. 
To further demonstrate the stability of the new asymmetric 
symbiotic pairs in panels (b)-(d) the BdG spectra are shown for different values of the inter-species 
interactions below the associated critical point $g^{(2)}_{12_{cr}}=0.81$. 
Two anomalous modes appear in the linearization spectra of these asymmetric states illustrated with red stars.
The respective stationary wave profiles are depicted in panels (e)-(g), where the dark (bright) soliton 
wavefunction is shown with dashed black (solid green) line. 
We observe that upon decreasing $g_{12}$ the asymmetry between the bright solitons increases,
as was also the case for the respective asymmetric but unstable states found in the immiscible regime.
Furthermore, for $g_{12}<0.7$ a background gradually builds up
for the bright solitons as is evident in the stationary
state shown in panel (e) of Figure~\ref{fig5}, revealing the miscible
character of the regime supporting these states. 
Our detailed BdG analysis indicates that for all values within the miscible regime that we have checked, 
the asymmetric states exist as stable 
configurations and should remain dynamically robust. This result is verified and highlighted in panels (h)-(m) 
of Figure~\ref{fig5} where we use as initial condition the stationary asymmetric states depicted in panels (e)-(g). 
We note that in all three cases panels (h)-(j) [(k)-(m)] show the evolution of the dark (bright) soliton component. 
Having studied both the static and the dynamical properties of the new asymmetric structures which bifurcate from the 
OP dark-bright states, we now turn our attention to the dynamics of
the anti-symmetric DB waveforms.
\begin{figure}[tbp]
\centering
\includegraphics[scale=0.4]{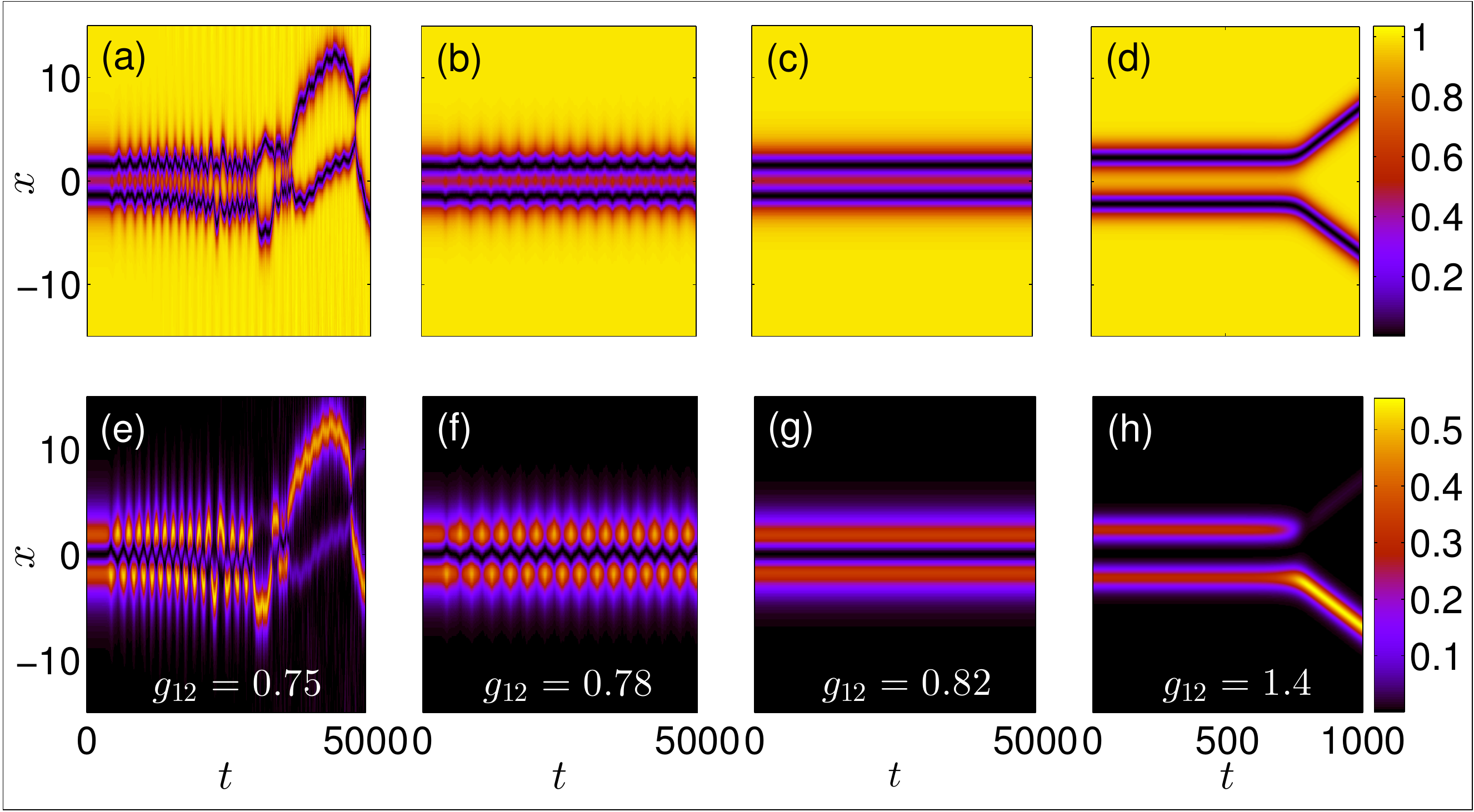}
\caption{Space-time evolution of the anti-symmetric stationary two-DB state for different values of the inter-species 
interaction coefficient $g_{12}$. From left to right the interaction is increased from $g_{12}=0.75$ to 
$g_{12}=0.82 > g_{{12}_{cr}}^{(2)}$, and then to $g_{12}=1.4 >g_{{12}_{cr}}^{(1)}$. Panels (\textbf{a})-(\textbf{d}) 
[(\textbf{e})-(\textbf{h})] 
correspond to the densities $|\psi_d|^2$ ($|\psi_b|^2$) of the dark (bright) component.}
\label{fig6}
\end{figure}   

In particular, we explore the long time evolution
of the OP DB states so as to reveal the  
decay mechanisms that such a pair suffers from.
The dynamics at different values of $g_{12}$ are summarized in Figure~\ref{fig6}, all initiated at equilibrium.
As before, upper row of panels (a)-(d) shows the spatio-temporal evolution of the dark soliton counterpart,
while panels (e)-(h) depict the corresponding bright component.   
From the stability analysis presented above it is expected that for all values 
of the interspecies interaction coefficient $g^{(2)}_{12_{cr}} < g_{12} < g^{(1)}_{12_{cr}}$
the anti-symmetric two dark-bright soliton states exist as stable configurations, and as such should be robust throughout the 
propagation.
This latter result is confirmed in panels (c) and (g) of Figure~\ref{fig6}, for $g_{12}=0.82$ which is slightly above the 
lower critical point. 
However, and as anticipated, a very different picture is found below the critical
point $g^{(2)}_{12_{cr}}$ depicted in panels (b) and (f), (a) and (e) respectively.
Starting with the former, we observe that slightly below $g^{(2)}_{12_{cr}}$
the initial stationary state quickly decays and
in (b) and (f) we observe periodic tunneling of the bright component between the two dark solitons,
while the dark solitons are only relatively weakly affected here. 
It is worth mentioning that similar tunneling dynamics has been identified and interpreted in terms of a bosonic Josephson 
junction model in~\cite{et}, but with the crucial difference that the soliton pair was further supported by the restoring 
force of a harmonic trap in that work, while in our present setup there
is no external potential that would keep the dark solitons in place.
Remarkably, in this regime despite the mass exchange,
the bound soliton pair does not disintegrate.
Further decreasing $g_{12}$, in panels (a) and (e) the effects of the instability are more drastic. 
The time scale of the bright component oscillations decreases and the dark solitons  vibrate more strongly.
After some time of almost periodic oscillations, a more irregular type of motion sets in, with both dark solitons (one filled 
by most of the bright component,
the other almost empty) eventually moving towards positive $x$ while separating and recolliding in the process, suggesting 
still a kind of effective attractive interaction between them.
This decay mechanism deep in the miscible regime is to be contrasted to the unstable dynamics in the immiscible regime, above 
$g^{(1)}_{12_{cr}}$.
In panels (d) and (h) we show the time evolution of the anti-symmetric two-DB soliton state at $g_{12}=1.4$.
While initially almost no dynamics is visible in the densities, especially no
oscillations within the bright component,
on intermediate time scales a strong asymmetry in the bright filling builds up (see again \cite{ljpp} for a more detailed discussion) 
and subsequently the filled and the empty dark soliton split,
showing no sign of effective attraction. 
Notice that the above described decaying mechanism is rather similar to the one observed for the unstable
asymmetric states for $g_{12}>1$ (see panels (d) and (h) of Figure~\ref{fig4}). 
 
\section{Discussion and future challenges}

In the present contribution we investigated in detail the stability and dynamics of bound pairs of dark-bright symbiotic 
solitons, which arise as nonlinear matter wave excitations in mixtures of Bose-Einstein condensates featuring inter-atomic 
repulsion. 
In particular, we explored the scenario of differently weighted inter- and intra-species interaction coefficients, 
breaking the integrability of the relevant nonlinear Schrödinger model.  
It was argued by means of a recently proposed variational approach~\cite{ljpp} and shown numerically that upon departing from 
the integrable limit bound states of such symbiotic entities exist for anti-symmetric bright soliton counterparts,
the so called solitonic gluons. 
These anti-symmetric states were found to be robust within a bounded interval of the inter-species 
repulsion coefficient $g_{12}$,
limited by critical points both in the miscible and in the immiscible regime of the model, 
associated with a supercritical and a subcritical pitchfork bifurcation respectively. 
Below and above these boundaries, i.e. deep in the miscible and the immiscible regime, respectively, the anti-symmetric pair 
becomes unstable.
Long-time propagation revealed differences, but also common characteristics of the decay mechanisms in the two domains of 
instability.
Specifically, a striking common feature is the relevance of bright mass transfer between the two dark solitons.
In particular, in the
miscible domain, we identified new stationary asymmetric  states 
that bifurcate from the anti-symmetric ones in a supercritical pitchfork. The stability of the new asymmetric 
states was also dynamically confirmed. Moreover, it was shown that a further decrease of the interspecies repulsion, 
results in asymmetric states with the bright solitons living on top of
a finite BEC background, highlighting in this way the
miscible character of such bound pairs.
In contrast to the above picture, upon entering the immiscible regime, 
we had found in our recent work~\cite{ljpp} that the destabilization of the anti-symmetric dark-bright 
soliton pair is caused by a subcritical pitchfork bifurcation involving an
unstable stationary, but as in the corresponding miscible 
domain also mass-imbalanced dark-bright soliton pair mode.
In the present work, we further explored the range of existence and stability of this asymmetric branch, demonstrating its 
overall instability and its substantial deformation upon entering the miscible regime.

There are many directions that are worth considering in the future along the lines of this work.
In particular, by fixing the intra-species interactions $g_{22}=0.95$ in this work, we have not addressed the 
fate of the dark-bright soliton pair states when actually approaching the Manakov (integrable) limit, which would require 
$g_{12}=g_{22}=1$. 
In this respect it would be particularly interesting to see if the asymmetric state fully stabilizes upon restoring
integrability or maintains its weakly unstable nature. Towards this direction, and since in the integrable limit 
exact solutions are available~\cite{shepkiv,prinari}, one could link the anti-symmetric and asymmetric states obtained here 
with the exact families of dark-bright soliton solutions known in the integrable case. 
Establishing a possible connection of this type would furthermore
open a new direction of exploring and understanding the symbiotic soliton pairs,
since in such a case one could depart once more from the integrable limit, 
but having at hand exact analytical expressions for the two-soliton problem rather than the approximate ones 
constructed only from the single soliton solution, as used in our present work. Furthermore, and also in this direction,
in the integrable model 
exact closed form expressions exist not only for static symbiotic states like the ones considered here, but 
also for moving and scattering ones. Based on these, one could hope to get insights into the collisional 
dynamics of symbiotic entities at least in the vicinity of the integrable limit, paving the way of a detailed understanding of 
features like the breathing state formation observed here.
Studies along these lines are left for future work.

\acknowledgments{P. S. gratefully acknowledges financial support by the 
Deutsche Forschungsgemeinschaft (DFG)
in the framework of the grant SCHM 885/26-1.
P.G.K. gratefully acknowledges the
support of NSF-PHY-1602994, the
Alexander von Humboldt Foundation, the
Stavros Niarchos Foundation via the Greek Diaspora
Fellowship Program and the ERC under
FP7, Marie Curie Actions, People, International Research
Staff Exchange Scheme (IRSES-605096).}



\end{document}